\documentclass{PoS}

\title{Redshift, Time, Spectrum -- the most distant radio quasars with VLBI}

\ShortTitle{Redshift, Time, Spectrum -- the most distant radio quasars with VLBI}

\author{\speaker{S\'andor Frey}$^{,a}$, Leonid I. Gurvits$^{b,c}$, Zsolt Paragi$^{b}$, Krisztina \'E. Gab\'anyi$^{d}$\thanks{The EVN is a joint facility of European, Chinese, South African, and other radio astronomy institutes funded by their national research councils. This work was supported by the European Community's Seventh Framework Programme, Advanced Radio Astronomy in Europe, grant agreement no.\ 227290, and the Hungarian Scientific Research Fund (OTKA, grant no.\ K72515). We thank L\'aszl\'o Mosoni and D\'avid Cseh for their contribution to the VLBI studies reviewed here.}\\
\llap{$^a$} F\"OMI Satellite Geodetic Observatory, P.O. Box 585, H-1592 Budapest, Hungary\\ 
\llap{$^b$} Joint Institute for VLBI in Europe, Postbus 2, 7990 AA Dwingeloo, The Netherlands\\
\llap{$^c$} Department of Astrodynamics and Space Missions, Delft University of Technology, 2629 HS Delft, The Netherlands\\
\llap{$^d$} Konkoly Observatory, Research Centre for Astronomy and Earth Sciences, Hungarian Academy of Sciences, P.O. Box 67, H-1525 Budapest, Hungary\\
E-mail: \email{frey@sgo.fomi.hu}, \email{lgurvits@jive.nl}, \email{zparagi@jive.nl}, \email{gabanyi@konkoly.hu}}

\abstract{
The highest-redshift quasars are still rare and valuable objects for observational astrophysics and cosmology. They provide important constraints on the growth of the earliest supermassive black holes in the Universe, and information on the physical conditions in their environment. Among the nearly 60 quasars currently known at redshifts z>5.7, only a handful are ``strong'' emitters in radio continuum. These can be targets of sensitive high-resolution Very Long Baseline Interferometry (VLBI) observations to reveal their innermost structure, down to $\sim$10 pc linear scales. We review the results of our earlier European VLBI Network (EVN) experiments on three of the most distant radio quasars known to date, and give a preliminary report on the EVN detection of a fourth one. The results obtained so far suggest that we see really young active galactic nuclei -- not just in a cosmological sense but also in terms of their active life in radio.
}

\FullConference{Resolving The Sky - Radio Interferometry: Past, Present and Future \\
         April 17-20, 2012\\
         Manchester, UK}

\begin{document}

\section{Introduction}

\subsection{R for redshift -- and for radio}

Quasars at {\it redshift} $z$$\sim$6 have become spectroscopically identified for somewhat more than a decade \cite{Fan00,Fan01}, the first candidates being selected by their extremely red colours in the Sloan Digital Sky Survey (SDSS). These $i$-dropout objects, for which the absorption on the short-wavelength side of the Lyman-$\alpha$ emission line falls in the $i$ photometric band while the emission appears only at longer wavelengths, first in the $z$ band, are located in the approximate redshift range of 5.7$<$$z$$<$6.5. This technique was successfully applied for discovering the bulk of the $z$$\sim$6 quasars known to date, over 50 objects -- not only in the SDSS but also in the Canada--France High-z Quasar Survey (CFHQS) (see e.g. \cite{Will10} for a review). As of today, the redshift record holder among quasars is J1120+0641 at $z$=7.085 \cite{Mort11}. It was discovered in the United Kingdom Infrared Telescope (UKIRT) Infrared Deep Sky Survey (UKIDSS). The $z$-dropout objects like this one can have redshift as high as 7.5. The search for quasars even more distant than the current record moves from the optical to the near-infrared regime, e.g. with the Panoramic Survey Telescope and Rapid Response System (Pan-STARRS) where the first $i$-dropout quasar at $z$=5.73 has recently been discovered \cite{Morg12}, or in the Visible and Infrared Survey Telescope for Astronomy (VISTA) Kilo-degree Infrared Galaxy (VIKING) public survey \cite{Find12}.

Only four of the known $z$$>$5.7 quasars (in the order of their discovery: J0836+0054 \cite{Fan01} at $z$=5.77; J1427+3312 \cite{McGr06} at $z$=6.12; J1429+5447 \cite{Will10} at $z$=6.21; J2228+0110 \cite{Zeim11} at $z$=5.95) show detectable continuum {\it radio} emission. The total 1.4-GHz flux density of the first three quasars is just in the order of 1~mJy. The most recently found source, J2228+0110, the second $z$$\sim$6 quasar after J1427+3312 selected by its radio emission, is somewhat weaker. Although the sample is still small, the radio-loud ratio among the most distant known quasars ($\sim$7\%) is remarkably close to the 8\%$\pm$1\% found by matching the bright ($i$$<$18.5) SDSS quasars at any redshift with the radio detections in the 1.4-GHz Faint Images of the Radio Sky at Twenty-centimeters (FIRST) survey \cite{Ivez02}. The rare radio-emitting high-redshift quasars are particularly valuable, for a variety of reasons. The ultimate evidence for synchrotron jets produced by accretion of the surrounding material onto supermassive black holes (SMBHs) in active galactic nuclei (AGN) can be found in the radio by high-resolution Very Long Baseline Interferometry (VLBI) imaging observations. If the radio emission is compact on scales probed by VLBI, it should come from an AGN. Indeed, as we will review in Sect.~\ref{VLBI}, compact radio structures in all four known radio quasars at $z$$\sim$6 have successfully been detected with VLBI. Moreover, compact radio sources that existed at around the epoch of reionization could serve as ``beacons'', illuminating the intergalactic gas in their line of sight. This offers a good perspective to use them for studying the absorption spectrum of the neutral hydrogen with sensitive next-generation radio instruments like the Square Kilometer Array (SKA) \cite{Cari04}.
   
\subsection{T for time}

The lookback {\it time} is $\sim$12.5~Gyr for $z$=6, and $\sim$12.7~Gyr for $z$=7. (We assume a flat cosmological model with $H_{\rm{0}}=70$~km~s$^{-1}$~Mpc$^{-1}$, $\Omega_{\rm m}=0.3$, and $\Omega_{\Lambda}=0.7$ troughout this paper.) The existence of $z$$\sim$6 quasars proves that accreting SMBHs with masses up to $\sim$$10^9$~$M_{\odot}$ have already assembled within several hundred million years after the Big Bang. Observing the earliest quasars of the Universe can constrain models of their birth and early cosmological evolution, the growth of the central SMBHs of active galactic nuclei, and their link to the host galaxy evolution via feedback mechanisms. Intriguingly, many of the intrinsic properties observed in the infrared, optical, and X-ray wavebands make the highest-redshift quasars very similar to their lower-redshift cousins, suggesting that they are already ``evolved'' objects even within 1~Gyr after the beginning of the Universe. Thus there are observational efforts going on to identify the ``real firsts''. For example, based on the lack of the infrared emission originating from hot dust, two $z$$\sim$6 quasars in a sample of 21 seem less evolved, as the amount of hot dust in the quasar host may increase in parallel with the growth of the central SMBH \cite{Jian10}. As we will see in Sect.~\ref{general}, the results of our high-resolution interferometric observations of radio-emitting sources also point to young objects, at least in terms of their radio jet activity. 
  
\subsection{S for spectra}

In an attempt to offer yet another tribute to Prof. Richard T. Schilizzi (RTS), and as an addition to the ingenious title of this conference, we choose {\it spectra} to represent S. In the past, Richard participated with us in numerous VLBI studies of radio quasars known as the most distant ones that time, e.g. \cite{Gurv92,Gurv94,Frey97,Para99,Gurv00}. Another one of his major research interests was the study of young radio-loud AGN -- Gigahertz-Peaked Spectrum (GPS) and Compact Steep Spectrum (CSS) sources -- and their evolution, e.g. \cite{Snel00} and references therein. It is no surprise that in the highest-redshift Universe, the two topics eventually converge: the earliest radio AGN, right after their ignition should necessarily be young. Observations indicate that the spectral slope of the radio continuum is steep for the most distant quasars (Sect.~\ref{VLBI} and \ref{general}; Fig.~\ref{spectra}) in the observed $\sim$1--5~GHz frequency range, which corresponds to $\sim$10--40~GHz in the rest frame of the sources. According to a plausible model \cite{Falc04}, the high-redshift steep-spectrum objects may represent GPS sources at early cosmological epochs. The first generation of supermassive black holes could have had powerful jets that developed hot spots well inside their forming host galaxy, on linear scales of 0.1--10~kpc. Adopting the relation between the source size and the turnover frequency observed in GPS sources for our ``typical'' high-redshift quasars, the angular size of the smallest ($\sim$$100$~pc) of these early radio-jet objects would be in the order of 10 milli-arcseconds (mas), and the observed turnover frequency in their radio spectra would be around 500~MHz in the observer's frame \cite{Falc04}. This spectral turnover has not been detected yet, but the structural and limited spectral information available for the $z$$\sim$6 radio quasars known to date fit well in the picture.

\section{The highest-redshift radio quasars with VLBI}
\label{VLBI}

Here we briefly summarise our VLBI imaging results obtained for the $z$$\sim$6 radio quasars known to date. These results came from a series of experiments performed with the European VLBI Network (EVN) starting in 2002, shortly after the discovery of J0836+0054 \cite{Fan01}, the first quasar in this category. All the VLBI experiments were conducted in phase-reference mode, involving regular observations of nearby bright, compact reference radio sources. This technique allowed us to detect the weak target quasars, and to determine their astrometric position with mas-scale accuracy. For all but one quasar, VLBI observations were made at both 1.6~GHz and 5~GHz frequencies. For the weakest quasar in the sample that was discovered most recently (J2228+0110 \cite{Zeim11}), work is still in progress, and as of now, only 1.6-GHz EVN data have been collected. In the following subsections, we list the individual sources in the order of their discovery and of the date of their VLBI observations.

\subsection{J0836+0054}
J0836+0054 was found in the SDSS data \cite{Fan01} as the first quasar at $z$$>$5.7 with a radio counterpart in the FIRST survey catalogue \cite{Whit97}, with 1.4-GHz flux density $S_{1.4}$=1.11$\pm$0.15~mJy. Its accurate redshift was later measured as $z$=5.77 \cite{Ster03}. Our first experimental EVN observations were conducted on 2002 June 8 at 1.6~GHz. We found that essentially all radio emission comes from a compact but slightly resolved source within $\sim$10~mas angular extent which corresponds to $\sim$60~pc linear size at the distance of the quasar. We could rule out that the quasar's image is multiplied by strong gravitational lensing \cite{Frey03}. (It turned out later that it's true for the other $z$$\sim$6 quasars as well, in contrast to earlier predictions \cite{Wyit02}.) Upon the successful detection at 1.6~GHz, we initiated 5-GHz EVN observations of J0836+0054. The data from 2003 November 4 verified that the source is compact ($<$40~pc) with a flux density $S_{5}$=0.34~mJy. Thus the spectrum of the source is steep; the variablity as a cause of the difference in flux densities is excluded by lower-resolution Very Large Array (VLA) observations performed nearly at the same time \cite{Frey05}. The two spectral points as a function of the rest-frame frequency are plotted in Fig.~\ref{spectra}, along with the measurements for the other $z$$\sim$6 radio quasars and a low-redshift object for comparison.  

\subsection{J1427+3312}
J1427+3312 ($z$=6.12) was identified as the first radio quasar above redshift 6 \cite{McGr06,Ster07}. Our 1.6-GHz and 5-GHz EVN imaging observations were conducted on 2007 March 11 and 2007 March 3, respectively. The source was clearly detected at both frequencies. Quite remarkably, there are two distinct radio components seen in the 1.6-GHz image of J1427+3312, separated by 28.3~mas, corresponding to a projected linear distance of $\sim$160~pc \cite{Frey08}. A similar result was published from an independent 1.4-GHz experiment conducted with the US Very Long Baseline Array (VLBA) \cite{Momj08}. Both radio components with sub-mJy flux densities appear resolved. 5-GHz radio emission on mas-scale was only detected for the brighter of the two, indicating again a steep radio spectrum (Fig.~\ref{spectra}), which is presumably the case for the other component which was too weak to be detected at the higher frequency. The double structure, the steep spectrum, and the separation of the components remind us to the Compact Symmetric Objects (CSOs), extremely young radio sources known in the more nearby Universe \cite{Wilk94,Owsi98}. If this analogy holds, the kinematic age of J1427+3312 could be in the order of 10$^3$ years. The motion of the components could in principle be detected and the expansion speed measured with repeated VLBI imaging in the future. The nature however is not very cooperative in this case: due to the time dilation caused by the extremely large cosmological redshift, the expansion would appear very slow, and one must wait at least for an astronomer's lifetime between the subsequent epochs of such a monitoring experiment.    

\subsection{J1429+5447}
J1429+5447 ($z$=6.21) is the most distant radio quasar known to date, found in the CFHQS \cite{Will10}. Our EVN images made on 2010 June 8 (at 1.6~GHz) and 2010 May 27 (at 5~GHz) show compact but somewhat resolved structures in the case of this source as well \cite{Frey11}. The steep radio spectrum of the VLBI-detected quasar (Fig.~\ref{spectra}) is similar to that of the previous two sources which have dual-frequency VLBI data available.

\subsection{J2228+0110}
J2228+0110 ($z$=5.95) was found by matching the optical detections of the deep SDSS Stripe 82 with the radio sources detected in the 1.4-GHz VLA A-array survey covering the same area \cite{Zeim11}. This quasar is different from the previous three in the sense that it falls below the detection threshold of the FIRST survey. Its peak brightness is 0.31~mJy/beam in the VLA Stripe 82 survey catalogue \cite{Hodg11}. A cautious approach to the VLBI detection led us to initiate 1.6-GHz observations first. The EVN experiment was conducted on 2011 November 1. The analysis of the data has not been completed yet, but according to preliminary results, J2228+0110 appears detected as a compact source, with a flux density much similar to that of the VLA one (L.I. Gurvits et al. 2012, in preparation).      

\section{Summary of the general properties}
\label{general}

\begin{figure}[!h]
\centering
\includegraphics[bb= 39 47 707 523, height=80mm, clip=]{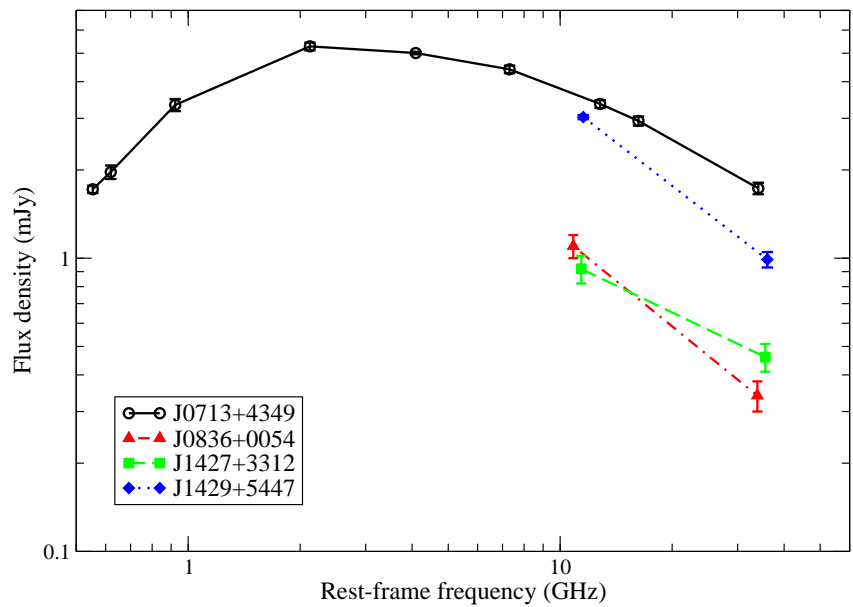}
\caption{Two-point radio spectra -- VLBI flux density versus frequency -- of the dominant components of 3 VLBI-imaged $z$$\sim$6 quasars. For comparison, spectral data points taken from the NASA/IPAC Extragalactic Database (NED) for J0713+4349 ($z$=0.52) are also plotted. This was the first CSO whose expansion speed could be measured with VLBI \cite{Owsi98}. The frequencies are shifted to the rest frame of each source. The total flux densities of J0713+4349 are scaled down by a factor of 0.0026, to match the luminosity distance if the source was moved to $z$=6. The lines connecting the data points serve to characterize the spectral shape only.} 
\label{spectra}
\end{figure}

By observing the sample of the four known $z$$\sim$6 radio quasars with the highest angular resolution provided by VLBI, we found that these are all compact sources. The bulk of their radio emission originates from regions well within 100~pc, clearly suggesting an AGN origin. The quasar J1427+3312 shows a double structure with components separated by about 160~pc, reminiscent of the structure of CSOs. It is possible that we see very young radio sources, like the GPS and CSS sources known in the less distant Universe. The measured moderate brightness temperatures ($\sim$10$^7$--10$^9$~K, substantially lower than the intrinsic equipartition limit, $\sim$5$\times$10$^{10}$~K, for powerful compact extragalactic radio sources \cite{Read94}) and the steep radio spectra in the rest-frame $\sim$10--40-GHz frequency range (Fig.~\ref{spectra}) can be considered as circumstantial evidence for the youth of these sources. The spectral indices are $\alpha$$\approx$$-0.6$...$-1.0$ for the three $z$$\sim$6 quasars where dual-frequency data are available. (The spectral index $\alpha$ is defined as $S\propto\nu^{\alpha}$, where $S$ is the flux density and $\nu$ the frequency.) 
In Fig.~\ref{spectra}, the broad-band spectrum of J0713+4349, a well-known CSO \cite{Owsi98} is also compiled from the total flux densities from the literature and plotted as a visual aid for comparison. The flux densities are scaled down to match the distance of the $z$$\sim$6 quasars. The spectral slope at the high-frequency end is quite similar to that of the three distant quasars. Obviously, additional lower-frequency observations would be needed to find the suspected spectral turnover for the $z$$\sim$6 objects -- a task very challenging with the current radio interferometric instruments due to the required spectral coverage, high sensitivity, and fine angular resolution.

A recent census of somewhat less distant VLBI-imaged radio quasars at $z>4.5$ \cite{Frey11} also suggests that the highest-redshift sample of compact radio sources is dominated by objects that do not resemble blazars that are characterised by highly Doppler-boosted, compact, flat-spectrum radio emission. Note that the continuum radio spectrum of bright blazars continues to be flat at much higher frequencies, in many cases up to several hundred GHz, e.g. \cite{Planck,Gere11}. If exist, blazar-type compact flat-spectrum AGN remain to be discovered at $z$$\sim$6. Certainly, the case of the extremely distant radio quasars is far from being closed, as new discoveries are expeced from on-going surveys, e.g. \cite{Morg12,Find12}, perhaps breaking the $z$=7 barrier soon.


\begin{thebibliography}{99}

\bibitem{Cari04}
Carilli C.L., Furlanetto S., Briggs F., et al. 2004, New Astron. Rev., 48, 1029

\bibitem{Falc04}
Falcke H., K\"ording E., Nagar N.M. 2004, New Astron. Rev., 48, 1157

\bibitem{Fan00}
Fan X., White R.L., Davis M., et al. 2000, AJ, 120, 1167

\bibitem{Fan01}
Fan X., Narayanan V.K., Lupton R.H., et al. 2001, AJ, 122, 2833

\bibitem{Find12}
Findlay J.R., Sutherland W.J., Venemans B.P., et al. 2012, MNRAS, 419, 3354

\bibitem{Frey97}
Frey S., Gurvits L.I., Kellermann K.I., Schilizzi R.T., Pauliny-Toth I.I.K. 1997, A\&A, 325, 511

\bibitem{Frey03}
Frey S., Mosoni L., Paragi Z., Gurvits L.I. 2003, MNRAS, 343, L20

\bibitem{Frey05}
Frey S., Paragi Z., Mosoni L., Gurvits L.I. 2005, A\&A, 436, L13

\bibitem{Frey08}
Frey S., Gurvits L.I., Paragi Z., Gab\'anyi K.\'E. 2008, A\&A, 848, L39

\bibitem{Frey10}
Frey S., Paragi Z., Gurvits L.I., Cseh D., Gab\'anyi K.\'E. 2010, A\&A, 524, A83 

\bibitem{Frey11}
Frey S., Paragi Z., Gurvits L.I., Gab\'anyi K.\'E., Cseh D. 2011, A\&A, 531, L5
 
\bibitem{Gere11}
Ger\'eb K., Frey S. 2011, Adv. Space Res., 48, 334 

\bibitem{Gurv92}
Gurvits L.I., Kardashev N.S., Popov M.V., et al. 1992, A\&A, 260, 82

\bibitem{Gurv94}
Gurvits L.I., Schilizzi R.T., Barthel P.D., et al. 1994, A\&A, 291, 737 

\bibitem{Gurv00}
Gurvits L.I., Frey S., Schilizzi R.T., et al. 2000, Adv. Space Res., 26, 719 

\bibitem{Hodg11}	
Hodge J.A., Becker R.H., White R.L., Richards G.T., Zeimann G.R. 2011, AJ, 142, 3

\bibitem{Ivez02}
Ivezi\'c \v{Z}., Menou K., Knapp G.R., et al. 2002, AJ, 124, 2364

\bibitem{Jian10}
Jiang L., Fan X., Brandt W.N., et al. 2010, Nature, 464, 380

\bibitem{McGr06}
McGreer I.D., Becker R.H., Helfand D.J., White R.L. 2006, ApJ, 652, 157

\bibitem{Momj08}
Momjian E., Carilli C.L., McGreer I.D. 2008, AJ, 136, 344

\bibitem{Morg12}	
Morganson E., De Rosa G., Decarli R., et al. 2012, AJ, 143, 142

\bibitem{Mort11}	
Mortlock D.J., Warren S.J., Venemans B.P., et al. 2011, Nature, 474, 616

\bibitem{Owsi98}
Owsianik I., Conway J.E. 1998, A\&A, 337, 69

\bibitem{Para99}
Paragi Z., Frey S., Gurvits L.I., et al. 1999, A\&A, 344, 51

\bibitem{Planck}
Planck Collaboration, Aatrokoski A., et al. 2011, A\&A, 536, A15

\bibitem{Read94}	
Readhead A.C.S. 1994, ApJ, 426, 51

\bibitem{Snel00}
Snellen I.A.G., Schilizzi R.T., Miley G.K., et al. 2000, MNRAS, 319, 445

\bibitem{Ster03}
Stern D., Hall P.B., Barrientos L.F., et al. 2003, ApJ, 596, L39

\bibitem{Ster07}
Stern D., Kirkpatrick J.D., Allen L.E., et al. 2007, ApJ, 63, 677

\bibitem{Whit97}
White R.L., Becker R.H., Helfand D.J., Gregg M.D. 1997, ApJ, 475, 479

\bibitem{Wilk94}
Wilkinson P.N., Polatidis A.G., Readhead A.C.S., Xu W., Pearson T.J. 1994, ApJ, 432, L87

\bibitem{Will10}
Willott C.J., Delorme P., Reyl\'e C., et al. 2010, AJ, 139, 906

\bibitem{Wyit02}
Wyithe J.S.B., Loeb A. 2002, Nature, 417, 923  

\bibitem{Zeim11}
Zeimann G.R., White R.L., Becker R.H., et al. 2011, ApJ, 736, 57

\end{thebibliography}
\end{document}